**R. Yu. Chaplynskyi[1,*], F. A. Danevich[1,2], D. V. Kasperovych[1], V. R. Klavdiienko[1], V. V. Kobychev[1], E. E. Petrosian[1], A. R. Podviianiuk[1], R. B. Podviianiuk[3], O. G. Polischuk[1]**

[1] *Institute for Nuclear Research, National Academy of Sciences of Ukraine, Kyiv, Ukraine*
[2] *Institute of Experimental and Applied Physics, Czech Technical University, Prague, Czech Republic*
[3] *Department of Physics, Syracuse University, Syracuse, USA*

*Corresponding author: chaplinskiyryu@gmail.com

# GAMMA SPECTROMETRY WITH CsI(Tl), NaI(Tl) AND $CdWO_4$ SCINTILLATION CRYSTALS USING A SILICON PHOTOMULTIPLIER

This study investigated using of silicon photomultiplier (SiPM) for scintillation γ-spectrometry with $CdWO_4$, CsI(Tl), and NaI(Tl) crystal scintillators. At room temperature, CsI(Tl) crystal scintillator provides the best performance, while the achievable energy resolution is lower compared to that obtained with conventional photomultiplier tube (PMT) with green-enhanced photocathode. These findings highlight the potential of SiPMs as a compact and cost-effective alternative to PMTs in nuclear physics applications, particularly for light portable spectrometers, such as radiation monitoring systems based on small unmanned aerial vehicles.

## 1. Introduction

At present, the most commonly used photodetectors in scintillation γ-spectrometry are vacuum photomultiplier tubes (PMTs) [1]. In these devices, scintillation signals are collected and amplified through a cascade of dynodes via secondary electron emission. PMTs provide several advantages over alternative photodetectors (PIN photodiodes, avalanche photodiodes etc.), including high energy resolution, large sensitive area, reliability and long-term stability. They can be coupled with scintillators of various types, dimensions, and emission spectra, which ensures substantial flexibility in the design of scintillation detectors. Nevertheless, PMTs also exhibit inherent limitations, such as sensitivity to magnetic fields, the requirement for stable high-voltage supply, fragility, large dimensions and mass. These factors restrict their applicability in lightweight portable spectrometers, particularly in radiation monitoring systems mounted on small unmanned aerial vehicles.

Silicon photomultiplier [2–5] is a relatively recent type of solid-state photodetector that approaches the performance of conventional PMTs in many respects. A SiPM comprises an array of microcells (typically 10–50 μm in size), each functioning as an independent photodiode. A reverse bias exceeding the avalanche breakdown voltage is applied to every cell, thereby providing high internal gain. The combined output from all microcells constitutes a current pulse proportional to the number of photons absorbed by the device.

The main advantages of SiPMs are low operating voltage (20–70 V), compact size, high mechanical stability, sub-nanosecond response time, insensitivity to magnetic fields, and a broad operating temperature range [6–8]. These properties make them particularly suitable as photodetectors for compact scintillation systems intended for nuclear spectrometry in field applications (portable and mobile detectors deployed outdoors), including airborne and space-based applications.

However, SiPMs also exhibit certain limitations. These include relatively high dark current, a considerable dark count rate (spontaneous triggering of individual cells), and a pronounced temperature dependence of their characteristics. Such factors must be taken into account in the design of spectrometric detector assemblies based on these photodetectors.

A possibility to use various scintillation crystals (LSO, BGO, and CsI(Tl) crystals with volumes in the range of ~1 $cm^3$) coupled to 1×1 $mm^2$ SiPMs was investigated in the early study [6]. Despite employing specialized nuclear electronics, the reported results indicated poor performance of SiPMs in scintillation γ-spectrometry. By contrast, modern SiPMs demonstrate substantially improved characteristics compared to earlier devices, including reduced dark count rate, afterpulsing, and optical crosstalk between individual cells [9–11]. These advances make SiPMs increasingly promising as photodetectors for γ-spectrometry in nuclear physics research.

In the present work, we report studies of signal waveforms and γ-ray spectra obtained with $CdWO_4$, CsI(Tl), and NaI(Tl) scintillation crystals coupled using optical grease Dow Corning Q2-3067 to SiPM MICROFC-30035-SMT-TR device manufactured by Onsemi (USA) [9]. The SiPM has an active area of 3×3 $mm^2$, the number of microcells is 4774, and the microcell fill factor is 64 %. The maximum spectral sensitivity is at 425 nm (41 % of photon detection efficiency at 5 V of overvoltage, dropping to 10 % at wavelengths of 320 and 680 nm) [9]. The readout was performed using commercial electronics based on an 8-bit 25 MHz digital oscilloscope module VDS1022I [12] and a specially developed Python-based software.

## 2. Experiment and results

### 2.1. Current-voltage characteristic of the SiPM

The current-voltage characteristic of the reverse-biased SiPM MICROFC-30035-SMT-TR was measured using a Keithley 480 pA-meter and a Twintex TP-4303N laboratory power supply. As shown in Fig. 1, a sharp rise in current occurs at a reverse bias voltage of approximately 25 V, corresponding to the onset of avalanche breakdown. Further increase in the reverse bias voltage results in a significant growth of the SiPM gain, but is simultaneously accompanied by an increase in dark current and dark count rate [9]. In the present study, a reverse bias voltage of 28 V was applied, which provided sufficient gain while keeping the dark current and noise of the SiPM at its technical specification.

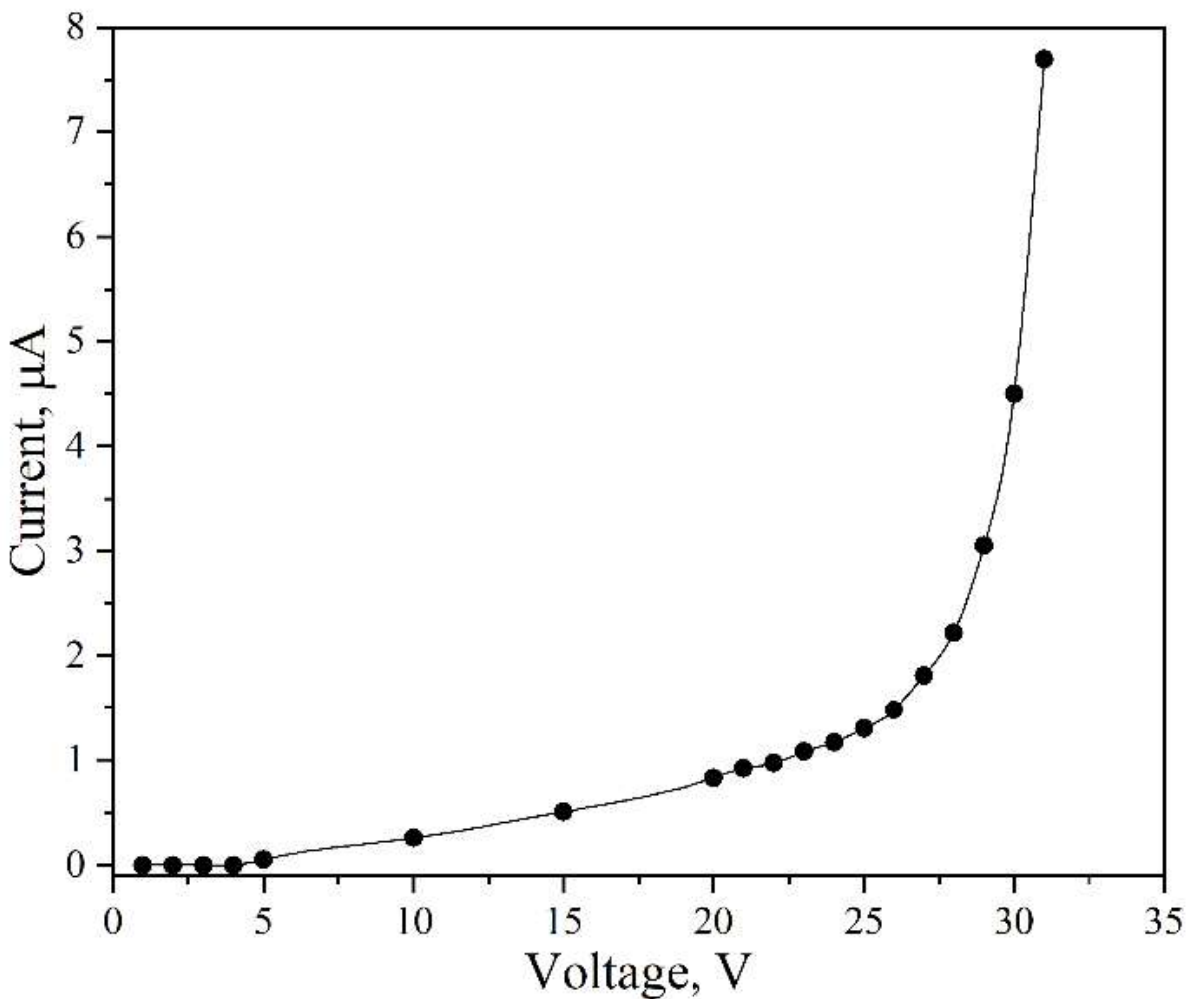

Fig. 1. Current-voltage characteristic of the SiPM MICROFC-30035-SMT-TR for reverse bias.

### 2.2. Characterization of scintillation detectors based on SiPMs

For the spectrometric measurements, the silicon photomultiplier was mounted on a printed circuit board equipped with a bias voltage filter and a load resistor (SMD 0603 package) installed on the bottom side. The photodetector board was placed in a polylactide support fabricated by fused deposition modelling 3D printing, which ensured stable positioning of the SiPM with respect to the crystal as well as reliable optical contact. The entire detector assembly was enclosed in a light-tight box.

Signal acquisition was carried out using an OWON VDS1022I oscilloscope module with a sampling rate of 100 MS/s in an experimental setup in which the input amplifier of the oscilloscope served as a preamplifier. The connection scheme is shown in Fig. 2.

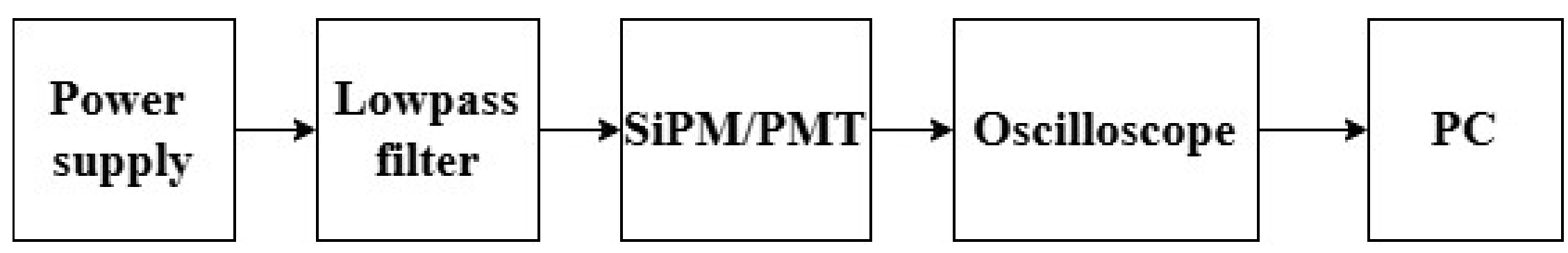

Fig. 2. Scheme of the experimental setup.

The scintillation crystals were coupled to the SiPM using optical contact and fixed with several layers of polytetrafluoroethylene (PTFE) tape. Characteristics of $CdWO_4$, CsI(Tl), and NaI(Tl) scintillation crystals used in the measurements are given in Table 1. Fig. 3 presents digitized signal waveforms recorded with the OWON VDS1022I oscilloscope module for each of the scintillators.

*Table 1.* **Properties of crystal scintillators used in the measurements (scintillation properties are compiled from [13–19])**

| Property | $CdWO_4$ | CsI(Tl) | NaI(Tl) |
|---|---|---|---|
| Scintillation decay time, μs | 13–15 | 0.8–1.0 | 0.23 |
| Wavelength at emission maximum, nm | 475–495 | 540–550 | 415 |
| Light yield, $\times 10^3$ photons/MeV | 6.2–28 | 52–66 | 38–41 |
| Crystal shape | Elliptical cylinder | Parallelepiped | Cylinder |
| Dimensions, mm | ∅(18–20)×20 | 10×6×6 | ∅25×25 |
| Encapsulation | PTFE film wrapped | Solid PTFE wrapped | Aluminium encapsulation |

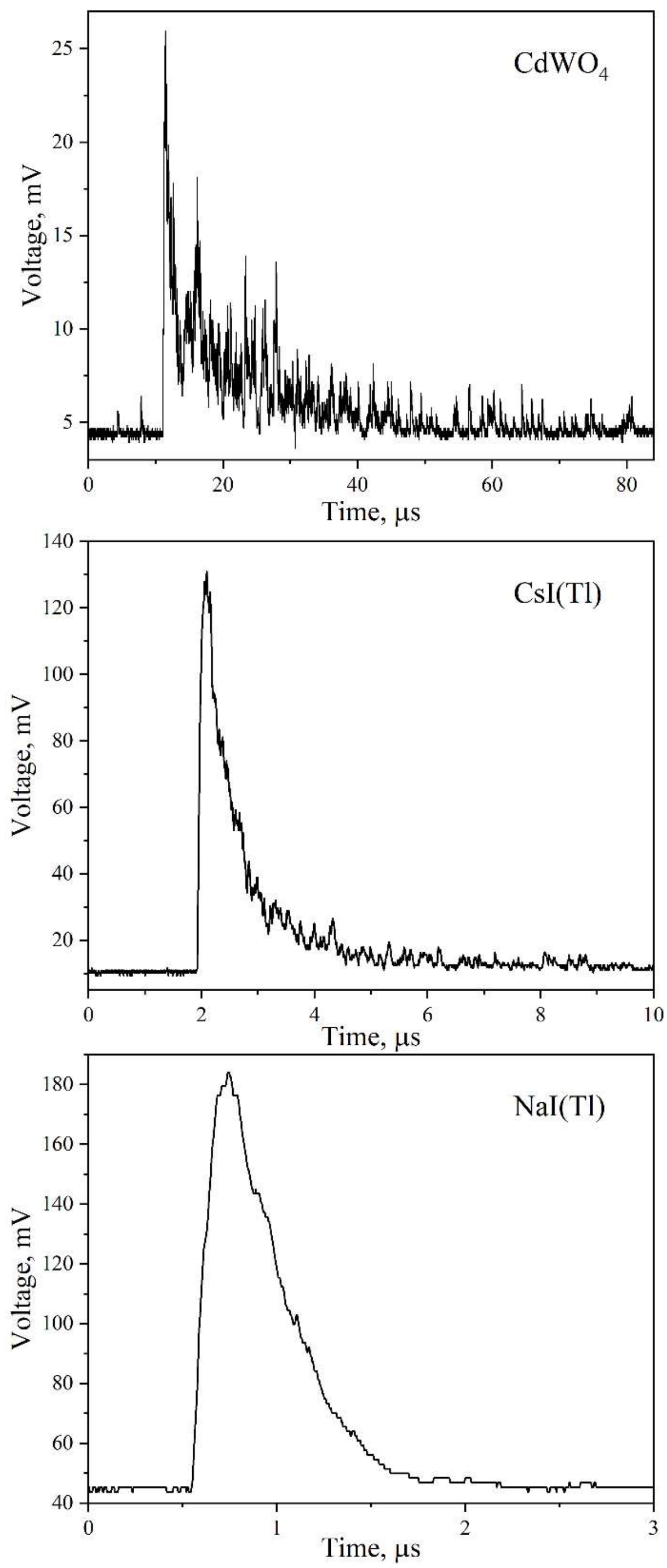

Fig. 3. Signal waveforms obtained with the SiPM and the $CdWO_4$ (*top panel*), CsI(Tl) (*middle panel*), and NaI(Tl) (*bottom panel*) scintillation crystals. Different time scales of the plots should be noted.

A distinctive feature observed in the $CdWO_4$ waveform is the separation of individual photoelectrons, manifested as discrete pulses. This effect results from a combination of factors: high time resolution of the SiPM, which allows the detection of individual photons; long scintillation decay time of cadmium tungstate; and its relatively low light yield.

The CsI(Tl) crystal, characterized by a high light yield and a favourable balance between light output and decay time, demonstrated the best compatibility with the investigated SiPM. These properties ensured the generation of a sufficient number of photons for reliable analogue signal formation, thereby enabling reasonably high spectrometric performance (see Table 1 and results below).

The energy spectra were measured using a $^{207}$Bi γ-ray source with the setup, in which the output signal was fed through a load resistor into the input of the oscilloscope module. The trigger level was set by the oscilloscope software as close as possible to the noise level. The digitized signal waveforms were transferred to a computer via a high-speed USB connection and stored on a hard drive for subsequent software processing. The energy spectra were constructed by integrating the area under the digitized waveforms over the entire data frame after baseline subtraction. The baseline was determined by averaging 100 ADC time bins at the beginning of each oscilloscope frame, ensuring accurate correction for electronic offset and noise prior to signal acquisition. For comparison of spectrometric performance, additional measurements were performed using Hamamatsu R6233 PMT (3” diameter, bialkali photocathode, spectral response window 300–650 nm) [20] under identical conditions (see Fig. 2).

The energy resolution was evaluated from the full width at half maximum (FWHM) of the γ-ray full energy absorption peaks. Among the studied samples, the best resolution was achieved with the CsI(Tl) crystal, yielding 11.6(4) % for the 570 keV peak and 9.1(5) % at 1064 keV. This performance can be attributed to the favourable matching of the scintillation emission spectrum with the SiPM photon detection efficiency and to efficient light collection enabled by the small crystal size, which provided a high signal-to-noise ratio. For comparison, the PMT measurements yielded FWHM values of 8.2(2) and 6.2(3) % for the 570 and 1064 keV peaks, respectively, representing better resolution than that obtained with the SiPM. The resulting spectra are shown in Fig. 4.

$CdWO_4$+PMT is known to provide good spectrometric performance [13], but for the $CdWO_4$+SiPM assembly, the relatively low light yield and long decay time resulted in signal amplitudes comparable to the dark count amplitude (as it is seen in the pre-trigger part of signal in Fig. 3, *top panel*). Analysis of the pre-trigger region indicates dark count rates of ~270 kHz, during the 15 μs $CdWO_4$ decay time, which significantly affects energy resolution. Consequently, dark count pulses significantly affected the integrated charge of the scintillation signals and distorted the amplitude spectrum, preventing the formation of resolvable peaks. Therefore, no pronounced full-energy peaks were observed in the spectrum from the present $CdWO_4$+SiPM assembly. However, we do not think that these features completely preclude the possibility of using these scintillators with silicon photomultipliers. The low temperature dependence of the light yield, chemical stability and non-hygroscopicity, high gamma-ray absorption efficiency, and the response to thermal neutron flux still leave cadmium tungstate in the group of scintillators of interest for mobile radiation monitoring. Improving the performance of gamma-spectrometry with cadmium tungstates and SiPMs remains an interesting challenge.

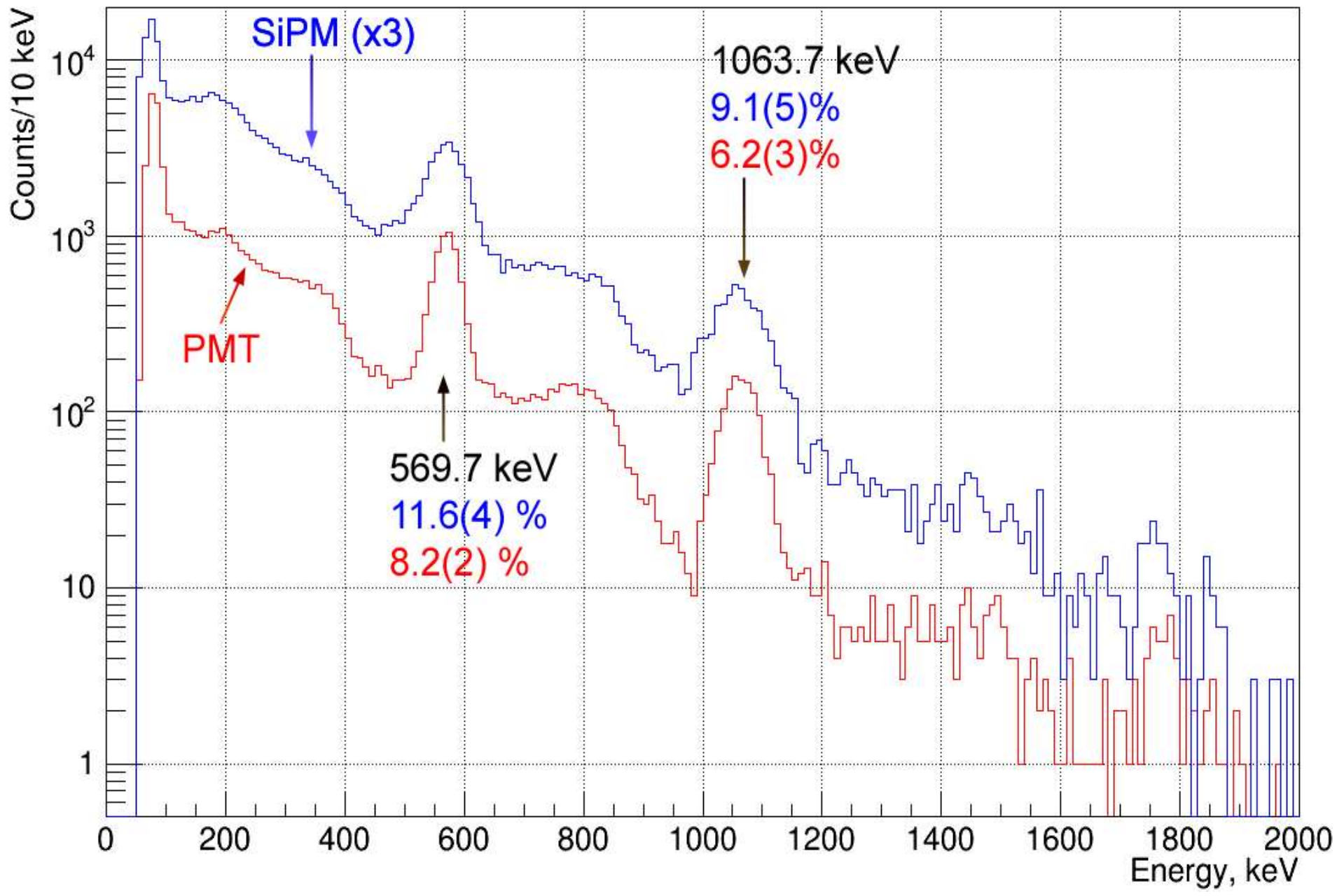

Fig. 4. Energy spectrum of γ-radiation from the $^{207}$Bi source obtained with a CsI(Tl) 10×6×6 mm$^3$ crystal coupled to the SiPM (*blue*) and the PMT (*red*). The FWHM values for the individual peaks are indicated in the corresponding colours. (See color Figure on the journal website.)

*Table 2*. **Obtained FWHM**

| Configuration | $^{207}$Bi 570 keV | $^{207}$Bi 1064 keV |
|---|---|---|
| CsI(Tl) + SiPM | 11.6(4) % | 9.1(5) % |
| NaI(Tl) + SiPM | 20.0(5) % | 15.0(9) % |
| $CdWO_4$ + SiPM | – | – |
| CsI(Tl) + PMT | 8.2(2) % | 6.2(3) % |
| NaI(Tl) + PMT | 8.0(2) % | 5.5(2) % |
| $CdWO_4$ + PMT | 8.5(2) % | 6.4(3) % |

For the NaI(Tl) crystal, despite its relatively high light yield and short decay time, the resolution was inferior to that of the CsI(Tl) assembly. This limitation arises primarily from the higher mismatch between the scintillator output window (∅25 mm) and the SiPM active area (3×3 mm$^2$), leading to lower light collection efficiency. For NaI(Tl), the measured FWHM values were ~20 % at 570 keV and ~15 % at 1064 keV. Results are summarized in Table 2.

## 4. Conclusions

Silicon photomultipliers, owing to their high gain, fast response, compact design, and low operating voltage, are becoming increasingly attractive for applications in nuclear physics. In the present study, it was demonstrated that scintillation γ-ray spectra can be obtained with a relatively simple and cost-effective setup consisting of an SiPM, a power supply, and a digital oscilloscope module.

Comparative measurements with $CdWO_4$, CsI(Tl) and NaI(Tl) crystals showed that the suitability of a scintillator for spectroscopy with SiPMs strongly depends on its light yield, light-collection conditions, and decay time. Due to the pronounced influence of dark counts, crystals with high light yield and relatively short decay time provided the most favourable performance, yielding resolvable γ peaks and the best energy resolution among the studied samples. In contrast, $CdWO_4$ with relatively wide range of reported $CdWO_4$ light yields (6.2–28 k photons/MeV, see Table 1) that reflects strong dependence on crystal quality and measurement conditions combined with the 15 µs decay time, leading to significant integration of dark counts into the energy signal, did not show reasonable spectrometric properties under given conditions, while NaI(Tl) demonstrated intermediate performance limited primarily by the geometric mismatch between the scintillator output window and the SiPM active area.

The energy resolution achieved with the SiPM-based assemblies was found to be ~40 % lower than that obtained with a conventional PMT under identical conditions. Nevertheless, the results confirm that modern SiPMs can serve a practical alternative to PMTs for scintillation γ spectrometry, particularly in applications requiring compact, low-voltage, light-weight and magnetically insensitive detectors. Further improvements in resolution may be expected through optimization of scintillator-to-SiPM coupling, reduction of dark count contributions, by cooling the detector assembly, and employing advanced signal processing techniques.

This work was supported by the National Academy of Sciences of Ukraine (Grant No. 0125U000172). The authors express their profound gratitude to the Armed Forces of Ukraine, whose defence of the country and its citizens against the invasion by the russian federation has made it possible to carry out the work reported herein. The authors are grateful to Anton Lymanets for his helpful discussions.

REFERENCES

1. G.F. Knoll. *Radiation Detection and Measurement* (Hoboken, NJ: John Wiley & Sons, 2010) 864 p.
2. P. Buzhan et al. Silicon photomultiplier and its possible applications. Nucl. Instrum. Methods A 504 (2003) 48.
3. D. Renker. Geiger-mode avalanche photodiodes, history, properties and problems. Nucl. Instrum. Methods A 567 (2006) 48.
4. R. Klanner. Characterisation of SiPMs. Nucl. Instrum. Methods A 926 (2019) 36.
5. M. Grodzicka-Kobylka, M. Moszyński, T. Szczęśniak. Silicon Photomultipliers in Scintillation Detectors for Nuclear Medicine. In: J.S. Iwanczyk, K. Iniewski (Eds.). *Radiation Detection Systems: Medical Imaging, Industrial Testing, and Security Applications*. 2nd ed. (Boca Raton: CRC Press, 2022) p. 225.
6. D.J. Herbert et al. First results of scintillator readout with silicon photomultiplier. IEEE Trans. Nucl. Sci. 53(1) (2006) 389.
7. D. Renker, E. Lorenz. Advances in solid state photon detectors. J. Instrum. 4 (2009) P04004.
8. M. Grodzicka et al. Energy resolution of small scintillation detectors with SiPM light readout. J. Instrum. 8 (2013) P02017.
9. Onsemi Silicon Photomultipliers (SiPM), Low-Noise, Blue-Sensitive. C-Series SiPM Sensors: Data Sheet.
10. Hamamatsu Multi-Pixel Photon Counter: Data Sheet.
11. S. Merzi et al. NUV-HD SiPMs with metal-filled trenches. J. Instrum. 18 (2023) P05040.
12. User Manual for VDS1022(I) PC oscilloscope.
13. L. Bardelli et al. Further study of $CdWO_4$ crystal scintillators as detectors for high sensitivity 2β experiments: Scintillation properties and pulse-shape discrimination. Nucl. Instrum. Methods A 569 (2006) 743.
14. M.J. Cieślak et al. Critical review of scintillating crystals for neutron detection. Crystals 9 (2019) 480.
15. C.L. Melcher. Perspectives on the future development of new scintillators. Nucl. Instrum. Methods A 537 (2005) 6.
16. M.J. Weber. Inorganic scintillators: today and tomorrow. J. Luminesc. 100 (2002) 35.
17. F.A. Danevich et al. α activity of natural tungsten isotopes. Phys. Rev. C 67 (2003) 014310.
18. C. Michail et al. Luminescence efficiency of cadmium tungstate ($CdWO_4$) single crystal for medical imaging applications. Crystals 10 (2020) 429.
19. Luxium Solutions. Scintillation Crystals.
20. Photomultiplier tube Hamamatsu R6233.